# Influence of HVOF parameters on HAp coating generation: An integrated approach using process maps

J.A. Hermann-Muñoz[1], J.A. Rincón-López[1], G.A. Clavijo-Mejía[1], A.L. Giraldo-Betancur[2], J.M. Alvarado-Orozco[3], A De Vizcaya-Ruiz[4], J. Muñoz-Saldaña[1, *].

[1] Centro de Investigación y de Estudios Avanzados del IPN, Unidad Querétaro, Libramiento Norponiente #2000, Fraccionamiento Real de Juriquilla, 76230 Santiago de Querétaro, México; jhermann@cinvestav.mx (J.A.H.-M.); jrincon@cinvestav.mx (J.A.R.-L.); gclavijo@cinvestav.mx (G.A.C.-M)

[2] CONACyT-Centro de Investigación y de Estudios Avanzados del IPN, Unidad Querétaro, Libramiento Norponiente #2000, Fraccionamiento Real de Juriquilla, 76230 Santiago de Querétaro, México; agiraldo@cinvestav.mx

[3] Centro de Ingeniería y Desarrollo Industrial, Av. Playa Pie de la Cuesta No. 702, Desarrollo San Pablo, 76125 Santiago de Querétaro, México; juan.alvarado@cidesi.edu.mx

[4] Centro de Investigación y de Estudios Avanzados del IPN, Unidad Zacatenco, Departamento de Toxicología, Av. Instituto Politécnico Nacional 2508, Col. San Pedro Zacatenco, Delegación Gustavo A. Madero, 07360 Ciudad de México, México; avizcaya@cinvestav.mx

**Abstract**

An experimental approach based on design of experiments, process maps and the analysis of deposition first stages to improve the biocompatibility of High-Velocity Oxygen Fuel (HVOF) hydroxyapatite (HAp) coatings is here presented. A two-level design of three factors ($2^3$) was performed using the stand-off distance (*SOD*), the fuel-oxygen ratio (*F/O*) and the powder feed rate (*PFR*). The effect of these experimental factors on the first stages of the coating formation was investigated to study the physical state of the particles before and after impacting the substrate. This study allowed the selection of the most suitable deposition parameter combinations to obtain HAp coatings with optimal crystallinity (> 45%), *Ca/P* ratio (approx. 1.67), phase content (> 95% of HAp), which guarantee the coatings mechanical stability and biocompatibility. The behavior of the coating within simulated body fluid (SBF) and cell culture (hFOB) was studied to analyze the apatite layer formation and the extracts cytotoxicity on human osteoblasts, respectively. The results show that the *F/O* ratio is the most influential factor on temperature and velocity on the in-flight particles and therefore on the coating properties. The SBF results confirmed the formation of an apatite layer after 14 days of immersion. Finally, the mitochondrial activity, measured by the MTS assay, and cell membrane integrity measures by LDH liberation assays, show that the coating released material does not induce toxicity on the exposed cells.

*Corresponding author: jmunoz@cinvestav.mx

1. **Introduction**

Thermal sprayed hydroxyapatite (HAp) coatings allow a natural fixation on metal implants for diverse applications, such as total hip and knee replacement that improve the patient biomechanics and extending the implants lifetime, and increasing patient's life quality [1]. Thermal plasma spray techniques are the only FDA approved processes to manufacture HAp coatings [2]. However, due to HAp instability at high temperatures (i.e., T > 1500 °C), this group of processes can lead to the formation of secondary phases, such as amorphous calcium phosphate (ACP), tricalcium phosphate (TCP), tetracalcium phosphate (TTCP), amongst others [3]. The implication of forming these phases is that under a physiological environment, they exhibit a higher solubility than HAp, compromising the coating long-term stability [4]. An alternative technique to obtain bioactive coatings with quality comparable or even higher than those obtained by plasma spray is the High Velocity Oxy-Fuel (HVOF) thermal spray [2,5]. In HVOF the powder reach higher velocities and lower temperatures allowing the formation of dense, crystalline and adherent coatings [6].

However, the properties of HVOF coatings are linked to the deposition parameters, which need to be optimized in order to control and analyze the effect of thermal spray parameters on coating performance. Among the several techniques to analyze and optimize HVOF thermally sprayed coatings, statistical methods such as design of experiments (DoE) are a useful alternative to predict responses related to the coating quality [7]. On the other hand, the use of in-flight particles diagnostic tools and the implementation of process maps have enabled a better understanding of the effect of different process variables on particle velocity and temperature in the plume [8–10].

The physical, chemical and kinetic conditions in which the particles reach the substrate, specially in the first stages have a great influence on the coating properties [11–13]. For instance, it is desirable that most of the particles upon impact with the substrate deform leading to the formation of disk shaped splats [12,14,15]. Further on, the phase content and adhesion of the coating are defined by the melt fraction and cooling rate of the particles after impacting the substrate [11,16]. As a result, the environmental stability, bioactivity and biocompatibility of the coatings are affected by the first stages of deposition and the overall build-up process [11,17].

To the authors' knowledge, there are no previous reports in the literature on the analysis of the thermal spray deposition process of high quality HAp-TiO$_2$ coating system by a combination of DoE and process maps using different diagnostic set-ups and statistical tools.

In this study, a systematic methodology to obtain HAp coatings on a TiO$_2$ bond coat by HVOF based on plume diagnostics, first stages of deposition, coating build up according to ISO 13779 and *in vitro* behavior is proposed. Design of experiments was thoroughly used and the coating formation was studied by the construction of process maps. To cover most of the critical variables and properties of both the coating and the spraying process, the study was divided in five stages in which the obtained characteristics were analyzed sequentially to select the most suitable combination of parameters to deposit crystalline, bioactive and biocompatible HAp coatings.

2. **Material and methods**

2.1. *Feedstock powder and substrates*

Commercial sintered HAp powder (Himed®) with nominal particle size between 20 and 63 μm were used in this study. The commercial powder was sieved using 345 and 500 mesh sizes (i.e., 25 μm < $x$ < 44 μm) to ensure a homogeneous particle size distribution. For splats deposition, 302 stainless steel (302 SS) of 2.5 x 2.5 cm coupons were polished to mirror finished quality. The substrates for coating deposition were cut in 1.2 x 1.2 cm squares and grit-blasted with alumina. Previous to HAp deposition, substrates were coated with a titanium oxide (TiO$_2$) bond-coat [18,19].

*2.2. Thermal spray process*

The splats and coatings were deposited in duplicate using a HVOF gun DJ2700 Hybrid (Sulzer-Metco) manipulated with a 6-axis robot arm (KUKA GmbH). Propane was used as fuel and nitrogen as carrier gas. The stand-off distance (*SOD*), fuel/oxygen ratio (*F/O*) and powder feed rate (*PFR*), known to have a great influence on the final coating properties, were chosen as inputs for a $2^3$ factorial design. A summary of parameters combinations is shown in Table I. The substrates for splats and coating deposition were heated at 300 °C before and during the powder projection [12].

*2.3. In-flight particles temperature and velocity*

The *in situ* monitoring DPV-evolution equipment (Tecnar) was used to measure the temperature, velocity, diameter and flow of the in-flight particles during HVOF coatings deposition.

*2.4. Powder size distribution*

Particle size distribution of HAp powder was measured in triplicate by a laser diffractometer (HELOS/BR, Sympatec GmbH), using the RODOS dry dispersion method with air pressurized at 0.2 bar.

*2.5. Powder, splats and coatings morphology*

Powder and splats morphology, as well as coatings microstructure were characterized using an environmental scanning electron microscope (ESEM-Philips XL30) at 10 kV electron acceleration voltage and a secondary electron detector. For cross section examination, coatings were mounted on epoxy, sectioned and polished to mirror finish. Additionally, characterization of splats was carried out using an optical microscope by image analysis with the Image J® software to quantify their covered area.

*2.6. Structural characteristics*

X-ray diffraction patterns (XRD) were recorded using a Rigaku DMax 2100 diffractometer, with CuKα radiation ($\lambda = 1.5406$ Å), between 20 to 70º on a 2θ scale with steps of 0.02º and counting time of 0.5 s. All samples were analyzed by Rietveld refinements of the XRD patterns to determine phase content, lattice parameters, crystallite size and *Ca/P* ratio following a methodology reported elsewhere [20]. Before and after SBF exposure, the structural changes on the coating surface were measured at 5º and 1º, respectively.

The structural characterization of powder and splats was complemented by Raman scattering measurements performed in a μ-Raman spectrometer (LabRam HR-800, Horiba Jobin Yvon) with a He-Ne laser at $\lambda = 632.8$ nm. To identify changes in intensity and displacement of HAp main bands, the Raman spectra were deconvoluted in the 400 - 4000 cm$^{-1}$ region with a Gaussian function using Origin®, obtaining an adjustment of $R^2 = 0.989$.

*2.7. In vitro analysis*

*2.7.1. Simulated Body Fluid (SBF) immersion*

Prior to biocompatibility assessment, the coatings fabricated with condition C6 were UV-sterilized for 12 h. Afterwards, HAp coating was immersed in SBF prepared according to [21]. The samples were soaked for periods of 3, 7, and 14 days in 30 mL of SBF solution at 36.5 ºC. After soaking, the samples washed out with distilled water and subsequently dried at room temperature.

2.7.2. *Cell culture*

A human fetal osteoblastic cell line denoted as hFOB 1.19 (American Type Culture Collection Number: CRL-11372) was used for cell viability and cytotoxicity assessment. Cells were cultured in Dulbecco's Modified Eagle Medium (DMEM) with 10 % fetal bovine serum and 1 % antibiotics, with 5 % vol. $CO_2$ in air at 37 ºC. The metabolic activity and cell cytotoxicity assays (MTS and LDH, respectively) were used to measure cell viability from exposure to the extracts, in compliance with ISO 10993-5 standard.

The extract method was performed to evaluate the cytotoxic potential of the products released by the coating and its chemical composition. According to ISO 10993-12 standard, an extraction ratio of 0.1 g/mL and an exposure time of 24 h at 37 ºC and 5 % vol. $CO_2$ in air were used. Thereafter, cells were seeded on 96 multi-well plates (cell density: $5 \times 10^3$ cells/cm$^2$) and exposed to dilutions with 0, 12.5, 25, 50, 75, and 100 % of the extract concentration for 72 h at 37 ºC and 5 % $CO_2$.

After the incubation period, cell viability was measured using the MTS (-3-(4,5-dimethylthiazol-2-yl)-5-(3-carboxymethoxyphenyl)-2-(4-sulphophenyl)-2H-tetrazolium) assay. The culture media was changed by 450 µL DMEM : 50 µL MTS reagent mixture and incubated for 4 h at 37 °C. Once completed the incubation time, the absorbance was measured at a wavelength of 490 nm, as an indicative of metabolically active cells. The highest absorbance of the control is considered as 100% and is used to compare with the tested conditions.

The lactate dehydrogenase (LDH) leakage was measured as an indicator of cytotoxicity induced by the coatings extracts. Cells treated with 5 % Triton-X100 for 30 min at 37 °C were used as a positive control of cytotoxicity. Thereafter, 50 µL of LDH assay (Roche Applied Sciences) were added to 50 µL of supernatant aliquots from each well and incubated for 30 min. Once completed the incubation time, the absorbance was measured at a wavelength of 490 nm, as an indicative of cell membrane damage. The highest absorbance of the control is considered as 0% and is used to compare with the tested conditions.

3. **Results and discussion**

3.1. *Feedstock powder*

It is well known that feedstock powder characteristics have an important effect on the final coating properties, since they influence the particles feeding conditions and the heat exchange both phenomena taking place inside the combustion chamber and in the plume, affecting coating crystallinity, phase purity, adhesion, amongst others [22,23].

The general characteristics of the feedstock powder used in this study are shown in Figure 1. The structural characterization by X ray diffraction (Figure 1a) was performed by comparing the obtained XRD pattern with the JCPDF 9-432 corresponding to the stoichiometric HAp phase according to the ISO 13779- 3 standard. No secondary phases were detected. After Rietveld refinement of the XRD pattern, the lattice parameters found were $a,b$ = 9.426 Å and $c$ = 6.893 Å. From the same analysis a *Ca/P* ratio of 1.66 was calculated considering atoms occupancies and multiplicities [20].

Further structural characterization of the powder was carried out by Raman spectroscopy. A typical HAp Raman spectrum is shown in Figure 1b. The $v1$, $v2$, $v3$ and $v4$ modes for $PO_4^{3-}$ were detected at 963, 400 – 490, 975 – 1150 and 560 – 620 cm$^{-1}$, respectively. Finally, the band exhibited at 3570 cm$^{-1}$ is attributed to the *OH*$^-$ group. The vibrational modes associated to the presence of other calcium phosphates phases were not detected.

A spherical morphology of the powder allows a good fluidity and therefore a continuous feeding of the powder to the gun [2]. The used powder shows a bimodal size distribution and a maximum frequency at 33 μm ($x_{10}$ = 1.49 ± 0.01 μm; $x_{50}$ = 27.63 ± 0.21 μm; $x_{90}$ = 43.42 ± 0.51 μm) as observed in Figure 1c. Figure 1d shows the feedstock powder as a combination of spherical agglomerates and fine particles.

It is also well known that the particle size distribution affects the heat transfer between the flame and the powder in which a narrow size distribution enhances the deposition efficiency [2,24]. Small particle sizes melt faster, promoting the formation of ACP and other phases. On the contrary, a wide distribution favors the heterogeneity of the particles melting states, generating greater porosity and roughness [23].

*3.2. Experimental space and in-flight particle conditions*

Among multiple variables involved in the HVOF process, the *SOD*, *PFR* and *F/O* ratio are reported to have the greatest influence on the residence time and heat transfer of the particle in the flame, thus affecting the main coating properties [7,23]. Therefore, the experimental space was limited with a $2^3$ factorial design as shown in Figure 2a. The upper and lower limits for each variable were stablished according to literature reports and previous works [18,25,26].

The DPV-evolution measurements of particle temperature and velocity for each one of the conditions established in the design of experiments were plotted in a diagram of *T* vs. *v*, portraying the first-order process map (Figure 2b). The process map method allows to understand the effect of deposition parameters on the energetic state of the particles. The *SOD* has a clear effect on particle velocity, the *F/O* ratio affects both temperature and velocity, while there is no a significant effect of *PFR* on the in-flight particles energy in the selected range. The reduction in particle velocity when *SOD* is increased can be related to the loss of drag force generated by the expansion of gases [27]. On the other hand, when the *F/O* ratio goes from 0.14 to 0.27 produces an increase in particle velocity and a drop in its temperature. These phenomena have been associated with the cooling of the flame due to fuel excess, which affects the chamber pressure and flame temperature (i.e. stoichiometric ratio for propane is about 0.21). The maximum flame temperature is reached at about 95% of the stoichiometric ratio with a significant decrease after this value [27–29]. In this regard, it is important to highlight two facts: firstly, the particle velocity was more sensitive to the variation of the process parameters, with values between 475 m/s and 675 m/s (42%). In contrast, the temperature values oscillated in less than 10%. Secondly, a greater effect of the *F/O* ratio on the velocity of the particle is observed at the upper limit of *SOD*. Similarly, a greater decrease in particle velocity is observed with the increment of *SOD* for flames with higher temperatures (F/O ratio = 0.14). Their effects on first stages of coating deposition will be discussed in the next section.

### 3.3. Initial deposition stages

The morphological and structural properties of the first particles deposited on the substrate are shown in Figure 3. Variations in morphology and particle distribution were observed by optical microscopy (Figure 3a). Two types of features were identified in the micrographs based on their morphology and contrast: a) a group of dark irregular agglomerates and b) light rounded deposits with litmus colors, can be observed. The nature of these deposits was studied using the SEM micrographs (Figure 3b), where darker zones correspond to particles impacting the substrate with a low melt-fraction leading themselves to thick agglomerates. In contrast, the litmus areas are particles impacting the substrate in a fully molten state allowing the formation of thin lamellae with different sizes and morphologies, matching with the general concept of what is known as splats in a thermal spray process [11,14]. The structural nature of these features was conducted by microRaman spectroscopy and results are shown in Figure 3c. In both zones, vibrational modes were identified in the region between 400 - 1500 cm$^{-1}$ due to the presence of the main bands of $PO_4^{3-}$, which are characteristic of calcium phosphates.

In the region between 400 - 900 cm$^{-1}$, the vibrational modes *v1* and *v2* of $PO_4^{3-}$ were detected for both, semi-molten and splats deposits, noticing significant differences in the full width at half maximum (FWHM) values. The higher value for the splats, indicates a loss of crystallinity. However, the broadening found for these bands is lower compared to those reported by other authors for HAp splats deposited by HVOF and plasma spray [30,31]. This is a positive result since amorphous phases exhibit higher solubility in a physiological environment [4]. The main HAp band corresponding to $v_1PO_4^{3-}$ was identified in the semi-molten particles with a shift of 10 cm$^{-1}$ compared to the feedstock powder, which has been previously associated to some amount of β-TCP phase [31]. In the splats case, the $v_1PO_4^{3-}$ peak position is in agreement with the expected value for the HAp. Within the 1000 – 1100 cm$^{-1}$ region, the vibrational modes of $v_3PO_4^{3-}$ were observed for both deposits. For splats, the featureless, broad and almost unobservable peaks between 1032 to 1081 cm$^{-1}$ on the curve are typical of ACP structure [13,31]. Finally, the presence of a broad peak at 1331 cm$^{-1}$ was detected for splats spectrum, suggesting the decomposition of HAp to produce $CaCO_3$ [31].

In summary, for semi-molten splats, 4 and 7 peaks were respectively identified and successfully deconvoluted. The peak position and FWHM values for each type of deposit were compared with those of the feedstock powder and functional groups assigned to each band (Table II).

To identify the relationship between the in-flight particle conditions and the splat formation, a quantification of the area occupied by the splats, also identified by the litmus area percentage was performed. The obtained data was plotted against particle velocities showing a linear relation between these two variables with $R^2$ adj = 0.925. i.e. faster particles generate a larger litmus area as can be appreciated in Figure 4. For a better understanding of this effect, a schematic representation of the changes observed in the morphology and size distribution of the first particles deposited on the substrate is shown in Figure 5. Despite observing an increase in the percentage of litmus area %, a decrease in the splats size was also detected when the velocity of the particles was increased. This phenomenon can be related to the residence time of the particle in the plume and the powder feedstock characteristics. The residence time is obtained by correlating the *SOD* with the measured particles velocity. Three types of morphologies were identified from the tested experimental conditions. In group 1, formed by C1 to C4 deposited at a *SOD* = 10 cm, a mixture of small rounded and elongated splats with sizes between 1 and 5 μm was observed. For these conditions, the average residence time of the particles in the flame was about 150 μs. As previously mentioned, the feedstock powder is a mixture between spherical agglomerates and fine particles. When this powder goes through the flame, a process of melting and droplet formation (agglomerate coalescence) occurs leading to splats formation upon impact [16,32]. Therefore, if the residence time of the particle in the flame is not enough to melt and complete the droplet formation, the agglomerated particle impacts the substrate forming structures, hereafter defined as sub-splats, which can content disk-like or splashing morphology. For group 2, composed by C7 and C8 (*SOD* = 20 cm and *F/O* = 0.27), fingered splats with sizes between 5 and 35 μm were observed. This occurs when the residence time of the particle in the flame is enough to melt and complete the droplet formation and the impact velocity of the particle is too high, generating instabilities during flattening and splat solidification leading to splashing [12,33]. Finally, in group 3, a homogeneous size distribution (about 30 μm) of disk-like splats was observed. Thus, disk-like formation splats occurs under critical conditions of melting, velocity and substrate temperature provided by the conditions deposited at 20 cm with a *F/O* ratio = 0.14 corresponding to C5 and C6 [12,14,33]. At this point the remaining question is, what is the origin of deceleration observed in Figures 4 and 5, which allows the formation of disk-like splats. This phenomenon is directly related to the torch length. It was observed a shorter flame for the conditions deposited with *F/O* ratio = 0.14, due to the amount of fuel [8]. As reported for Valarezo *et al.*, it is expected that longer flames reduce the resistance of the surrounding air due to the low density of the gases at high temperature. Thus, in shorter flames, an increase in the air resistance is observed leading to a higher deceleration. Specifically, in the case of conditions deposited with *SOD* = 20 cm and *F/O* ratio = 0.14, the substrate was located almost at the end of the flame indicating that the loss of drag force and the air resistance were maximal as evidenced by *in situ* measurements. All conditions were used to build up the coating and discussed in the next section.

3.4. *Coating characteristics*

The surface characteristics of the coatings deposited with a *PFR* = 16 g/min are compared to their properties during the initial stages of deposition as shown in Figure 6 a and b. A homogeneous surface is observed with the increase of *SOD*, while coating delamination occurs by increasing the *F/O* ratio. As mentioned before, an increase in splat size is obtained by increasing the *SOD*, leading to a homogeneous coating. On the other hand, higher splats density is observed with higher *F/O* ratio. The coatings deposited with these conditions showed delamination. The same behavior was observed for the conditions deposited with lower *PFR*. As is shown in Figure 6, conditions in which the particles impact the substrate at higher temperatures and lower velocities (C5 and C6) generating homogeneous coatings, which can also be related to the characteristics of deposited splats (disk-like splats). The coatings delamination (conditions deposited with *F/O* = 0.27 and *SOD* = 10 cm) can be explained, based on the combination of high velocity, low temperature and short residence time of the particles. These characteristics mainly generate non-melted and a low fraction of semi-molten particles, compromising the coating adhesion due to an increase in the compressive residual stresses (peening effect) [27]. In general, the characteristics of the deposited coatings confirm those observed in the initial deposition stages.

The XRD patterns of the deposited coatings are shown in Figure 7a. and compared with stoichiometric HAp phase (JCPDF 9-432 based on ISO 13779-3 standard). No secondary phases such as TCP, TTCP, among others were observed. Conditions C3 and C4 are not shown due to coating delamination.

The Rietveld refinement of XRD patterns was performed having goodness coefficients ($\chi 2$) close to 2 from which crystallite sizes and *Ca/P* ratios listed in Table III were obtained. The calculated *Ca/P* ratios are within the range (1.67 – 1.76) stablished by ISO 13779-2 standard, except for condition C5 (1.80). The determined structural characteristics by the aforementioned methods showed high quality coatings with promising properties for *in vitro* performance.

Typical SEM micrographs of the HAp/TiO$_2$ coating cross section deposited with C6 conditions are shown in Figure 7 b-c. A homogeneous and well-adhered HAp coating is observed over a thicker and denser TiO$_2$ bond-coat. The presence of some defects as cracks and holes on the cross section are related to the metallographic preparation.

*3.5. In vitro analysis*

Based on the previous results, the selected sample to perform *in vitro* assessments was the C6 condition, which was deposited with the following parameters: *SOD* = 20 cm, *PFR* = 16 g/min and *F/O* ratio = 0.14.

*3.5.1. SBF test*

To study the *in vitro* biomineralization process of the coating, blood plasma conditions were reproduced in SBF. The changes on the coating surface morphology after 0, 3, 7 and 14 days of exposure to SBF are shown in Figure 8 a-d. A granular morphology is observed for the coating before exposure and after 3 days in SBF. After 7 days of immersion in SBF a dune-like morphology is observed in some areas of the coating surface, whereas in 14 days these dunes cover the entire surface. The observed microstructure evolution is in agreement with literature reports, associated to the formation of a bone-like apatite layer of a bioactive material [34,35]. Data obtained from Rietveld refinement of XRD patterns allowed to quantify differences in the *Ca/P* ratio and crystallite size of the coating after 3, 7 and 14 days of SBF exposure (Figure 8e).

The behavior of the *Ca/P* ratio is as follows: A Ca-rich surface is observed after 3 days of exposure followed by a Ca-poor surface at 7 days and ending with a value very similar to the initial at 14 days, which indicates an ionic exchange between the coating and the physiological medium, as reported by Kim *et al.* [35]. For the crystallite size, a decrease is observed after 3 days of exposure indicating the first stages of formation of an amorphous layer that corresponds to bone-like apatite as already reported elsewhere. This process can be described as a dissolution/reprecipitation reaction process.

*3.5.2. Cytotoxicity*

Figure 9 shows the cell viability (a) and cytotoxicity (b) after 72 h of exposure at different extract concentrations of the condition C6 with and without $TiO_2$ bond-coat. No changes were observed at tested conditions, related to cell membrane lysis or cell death (LDH) and cell metabolic activity (MTS), after exposure to each extract concentration.

The statistical analysis showed differences between the coatings deposited with and without $TiO_2$ bond-coat in the metabolic activity for extracts concentrations of 75% and 100%, observing some loss of viability with the extracts from coatings deposited with $TiO_2$ bond-coat (Figure 9a). On the other hand, for the LDH assay, higher cytotoxicity was induced by the extract of 12.5% of the coating deposited without bond-coat compared to the same concentration for a coating with $TiO_2$ bond-coat. Despite of these specific differences, the statistical analysis showed that bond-coat has not effect on the cell viability or cytotoxicity.

4. **Conclusions**

The current study of the deposition process starting from the in-flight particles condition, initial stages characteristics, coating structural and microstructural properties to finally analyze its *in vitro* behavior, allowed to understand the influence of process parameters in each stage, and to select the most adequate conditions for biocompatibility testing. The main findings in this study are listed below:

- No significant influence of the PFR was observed neither in the in-flight particle conditions, splats formation nor in the coating properties.

- The combination of the shortest *SOD* (10 cm) and the highest *F/O* ratio (0.27) used parameters results in faster and colder particles with a short residence times in the flame, leading to coating delamination.

- A disk-like splat morphology was obtained using the following combination of parameters: *SOD* = 20 cm, *F/O* ratio = 0.14 and *PFR* = 16 g/min. The obtained coating showed a homogeneous surface, high crystallinity and *Ca/P* ratio that are within the range stablished by the ISO-13779 standard. The coating also showed a bioactive behavior.

- The material released by the coatings during the cytotoxicity assays does not induce a cytotoxic response.

**Acknowledgments:** The authors thank to CONACYT for the financial support. This project was funded by CONACYT Projects 272095, 279738 and 279738 (National Laboratory program) and carried out partially at CENAPROT, LIDTRA and LANIMFE national laboratories, LISMA and the Toxicology Department at Cinvestav - Zacatenco. The authors also thank Marisela Uribe-Ramirez, Jose Eleazar Urbina-Alvarez and Adair Jimenez-Nieto for technical support.

**Table I.** Combination of parameters for splats and coating deposition.

| Sample ID | Deposition parameters | | |
| --- | --- | --- | --- |
| | *F/O ratio* | *SOD* (cm) | *PFR* (g/min) |
| C1 | 0.14 | 10 | 8 |
| C2 | 0.14 | 10 | 16 |
| C3 | 0.27 | 10 | 8 |
| C4 | 0.27 | 10 | 16 |
| C5 | 0.14 | 20 | 8 |
| C6 | 0.14 | 20 | 16 |
| C7 | 0.27 | 20 | 8 |
| C8 | 0.27 | 20 | 16 |

**Table II.** Peaks position, FHWM and assigned bands determined for feedstock powder, splats and semi-molten particles. Reported values.

| Splats | | Semi-molten | | Feed stock powder | Reported band |
|---|---|---|---|---|---|
| Position Peaks (cm$^{-1}$) | FHWM | Position Peaks (cm$^{-1}$) | FHWM | Position Peaks (cm$^{-1}$) | |
| 430.2 | 69.9 | 456.5 | 40.3 | 427-442 | $v2\ PO_4^{3-}$ |
| 469.7 | 127.6 | | | | |
| 614.4 | 81.6 | 607.1 | 47.8 | 577-613 | $v4\ PO_4^{3-}$ |
| 678.0 | 62.8 | | | | |
| 962.5 | 37.1 | 975.5 | 21.0 | 960 | $v1\ PO_4^{3-}$ |
| 1043.9 | 279.9 | 1060.0 | 42.6 | 1026-1074 | $v3\ PO_4^{3-}$ |
| 1331.1 | 113.0 | | | | $CaCO_3$ |

**Table III.** Crystallite size and *Ca/P* ratio calculated from Rietveld Refinements

| Sample/Condition | Crystallite Size (Å) | Ca/P | $\chi^2$ |
|---|---|---|---|
| C1 (0.14;8;10) | 645.88 | 1.69 | 1.4 |
| C2 (0.14;16;10) | 508.60 | 1.72 | 1.6 |
| C3 (0.27;8;10) | | Delaminated | |
| C4 (0.27;16;10) | | Delaminated | |
| C5 (0.14;8;20) | 824.52 | 1.80 | 1.5 |
| C6 (0.14;16;20) | 716.99 | 1.69 | 1.4 |
| C7 (0.27;8;20) | 839.42 | 1.76 | 1.4 |
| C8 (0.27;16;20) | 758.04 | 1.69 | 1.5 |

**Figure Caption**

**Figure 1.** Characteristics of HAp commercial powder a) XRD pattern, b) FTIR spectrum, c) particle size distribution, that includes cumulative and distribution density, d) a typical SEM micrograph of the powder is also included.

**Figure 2.** Results of the process maps analysis based on a) current experimental space and b) Measurements obtained with the DPV evolution

**Figure 3.** Characteristics of the first stages of coating deposition recorded by a) optical microscopy b) SEM. Note, the three typical regions either of semimolten, disc shaped splats or spherical deposits. In c) Typical Raman spectra from the different deposited regions are shown.

**Figure 4.** Relationship between particle velocity and the %area covered by light rounded particles (%litmus area). $R^2$ ajus=0.925.

**Figure 5.** Variation of splats morphology and size distribution as a result of the in-flight particle conditions. The splats were identified according to their shapes as group 1) sub-splats, 2) splashing and 3) disk-like splats, as a function of the residence time of the particle in the flame.

**Figure 6.** Effects of the stand-off distance and Fuel/ Oxygen Ratio on the coating surface properties recorded by photographs.

**Figure 7.** XRD patterns of the coatings deposited with the conditions of the experimental space (a). Cross section of a typical coating deposited with C6 conditions and recorded at b) 250X and c) 500X.

**Figure 8.** Morphological changes on the surface as a result of the bioactivity evaluation for C6 coating after a) 0 , b) 3 c) 7 and d) 14 days of immersion in SBF. The variation of the *Ca/P* ratio and crystallite size obtained from the XRD Rietveld refinement after each period of exposure is shown in e).

**Figure 9.** Cell viability after 72h of exposure to different extract concentrations**.** a) MTS assay. b) Cytotoxicity by LDH assay. Data were compared (mean, SD, *n=*6) using a two-way ANOVA, (*) (**) represents the differences between HAp and HAp/TiO$_2$ coatings with a $p < 0.05$ and $p < 0.01$, respectively.

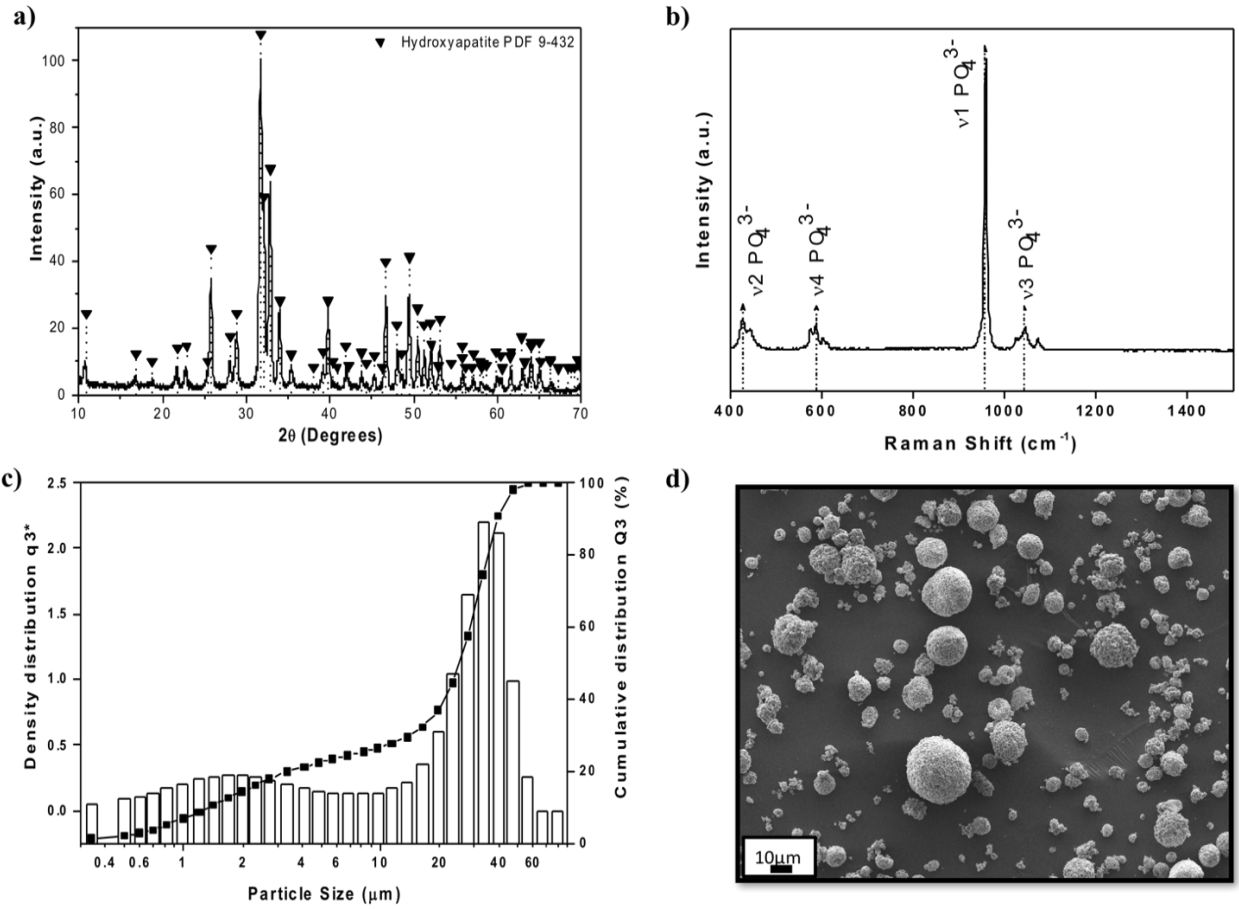

Figure 1.

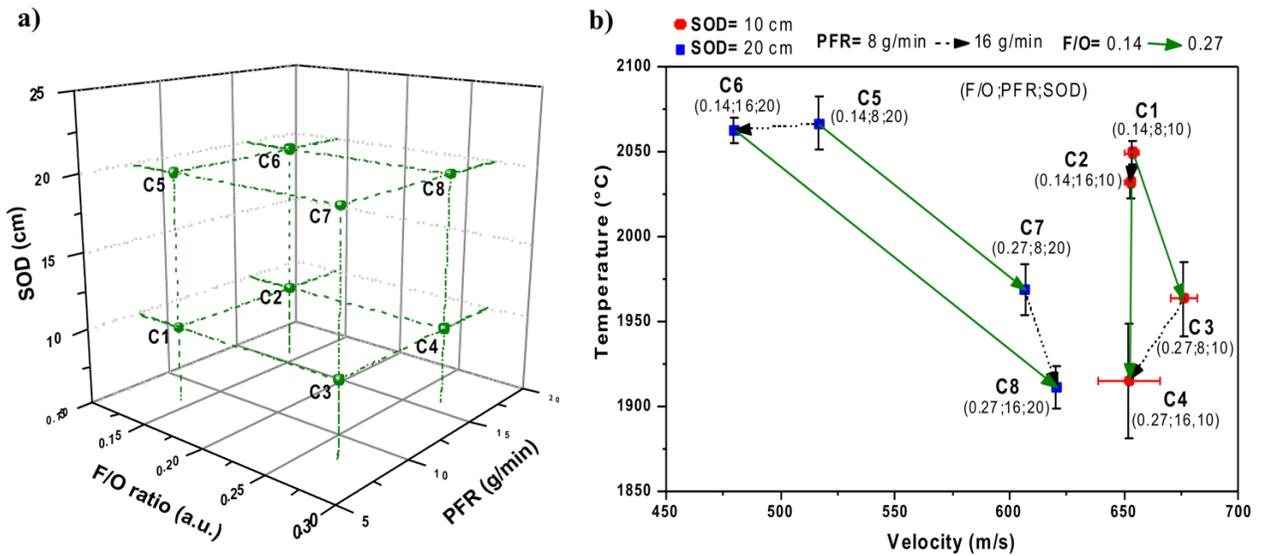

Figure 2.

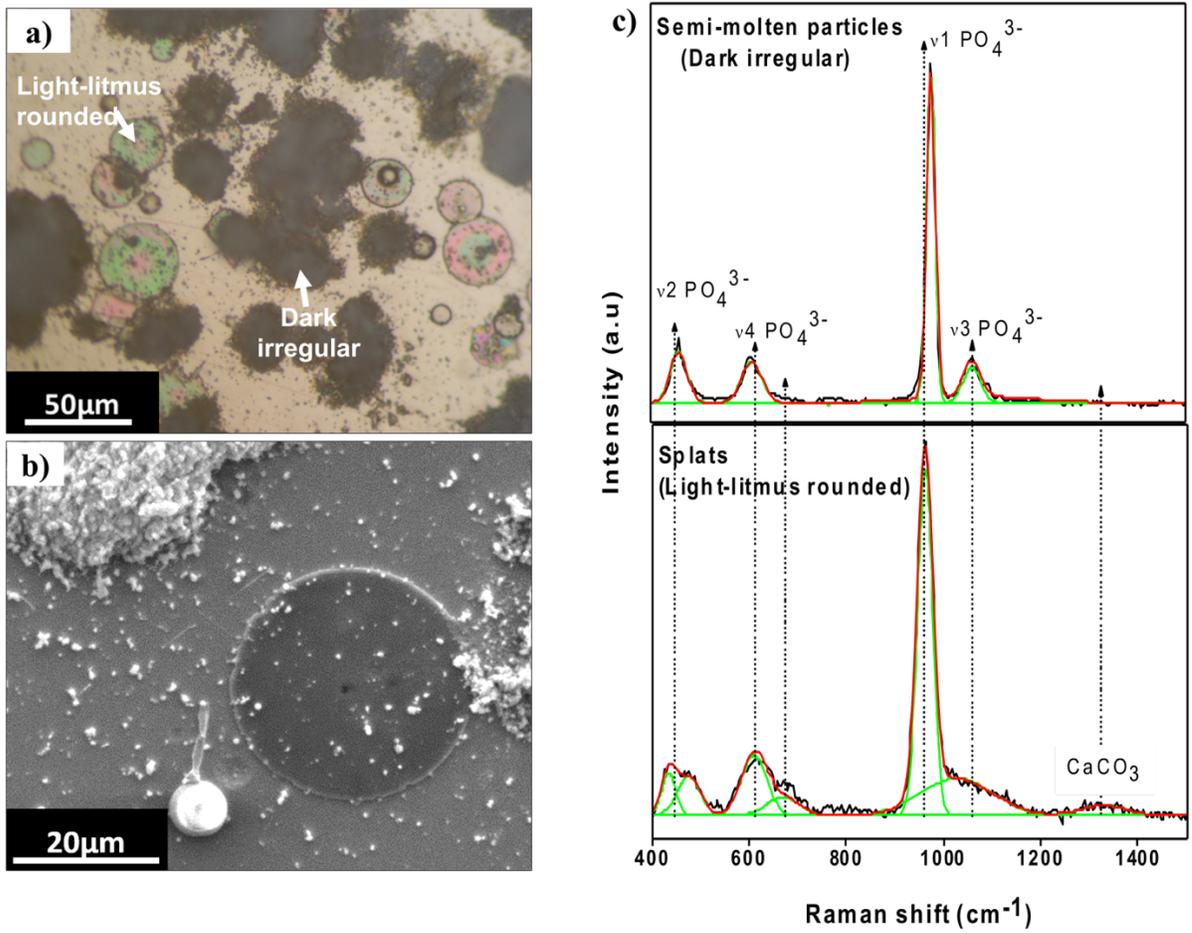

Figure 3.

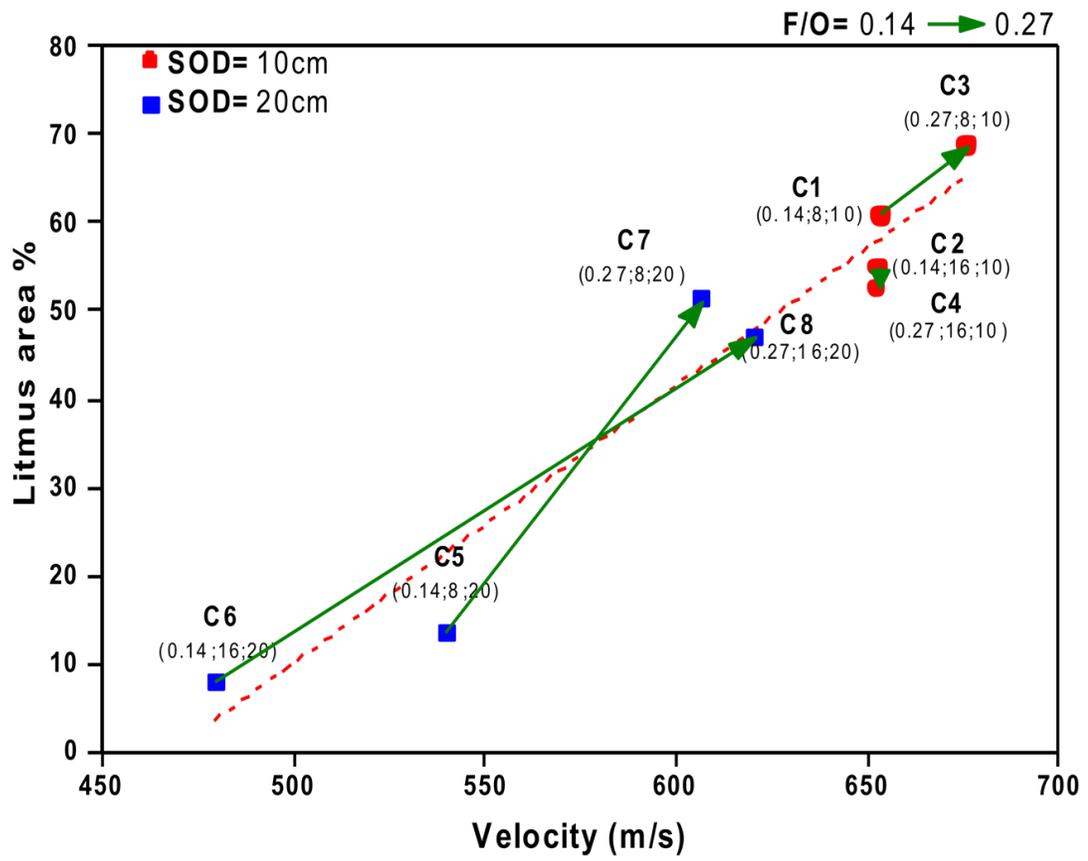

Figure 4.

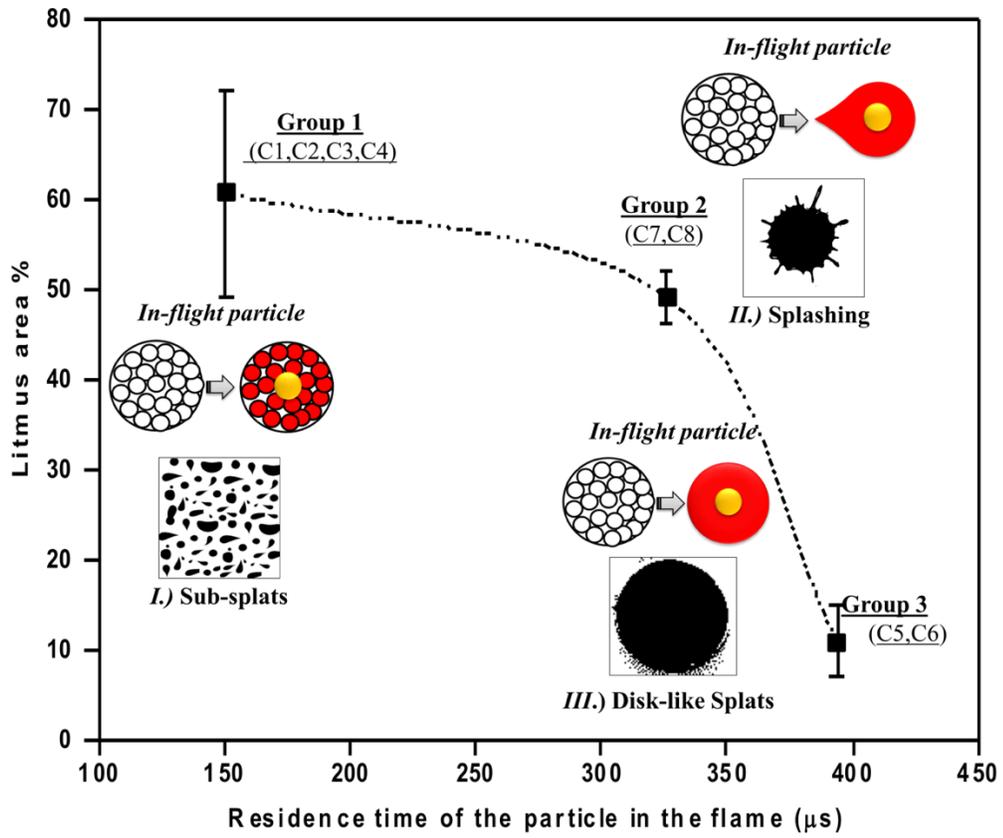

Figure 5.

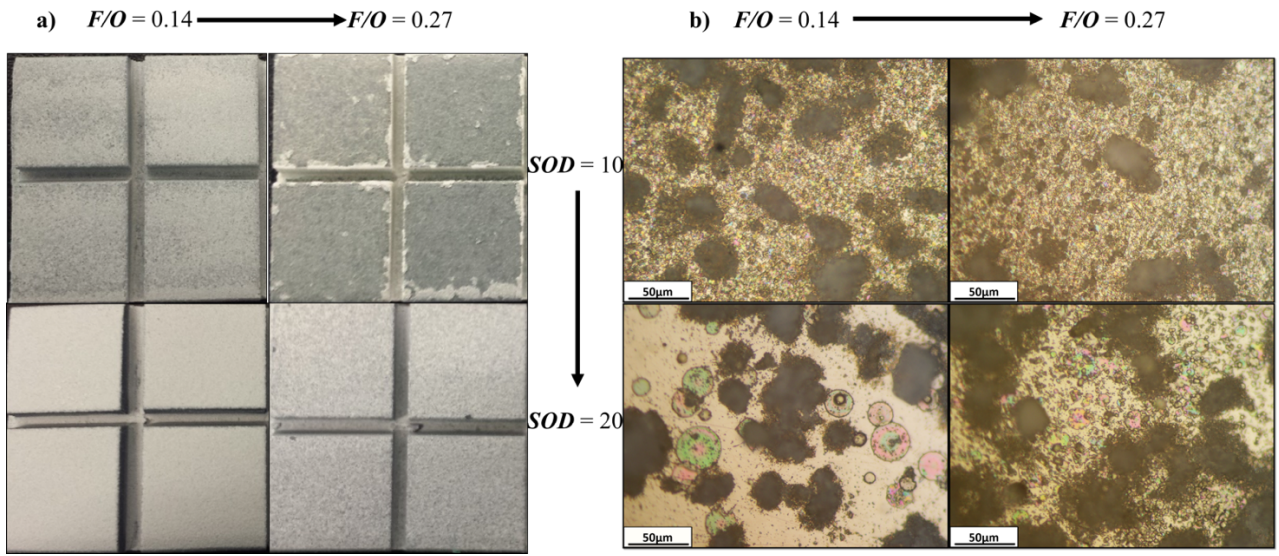

Figure 6.

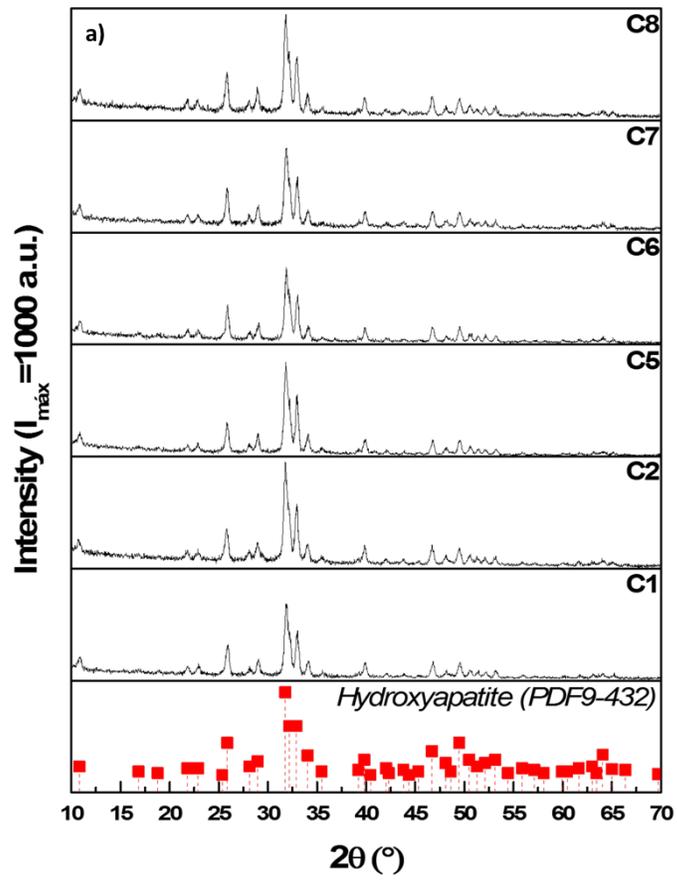
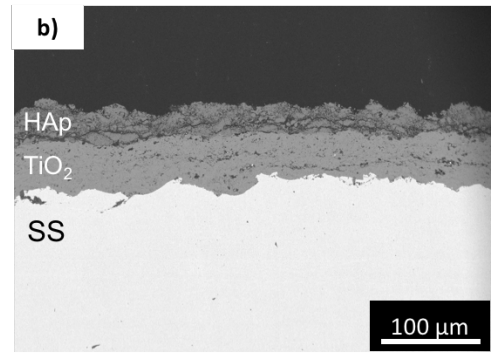
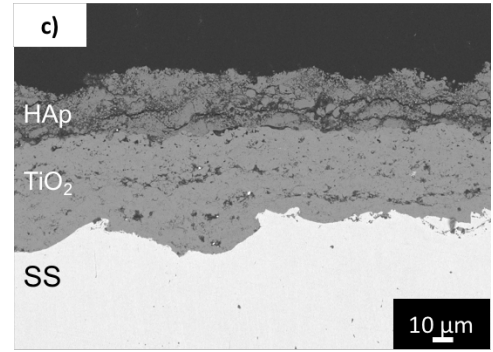

Figure 7.

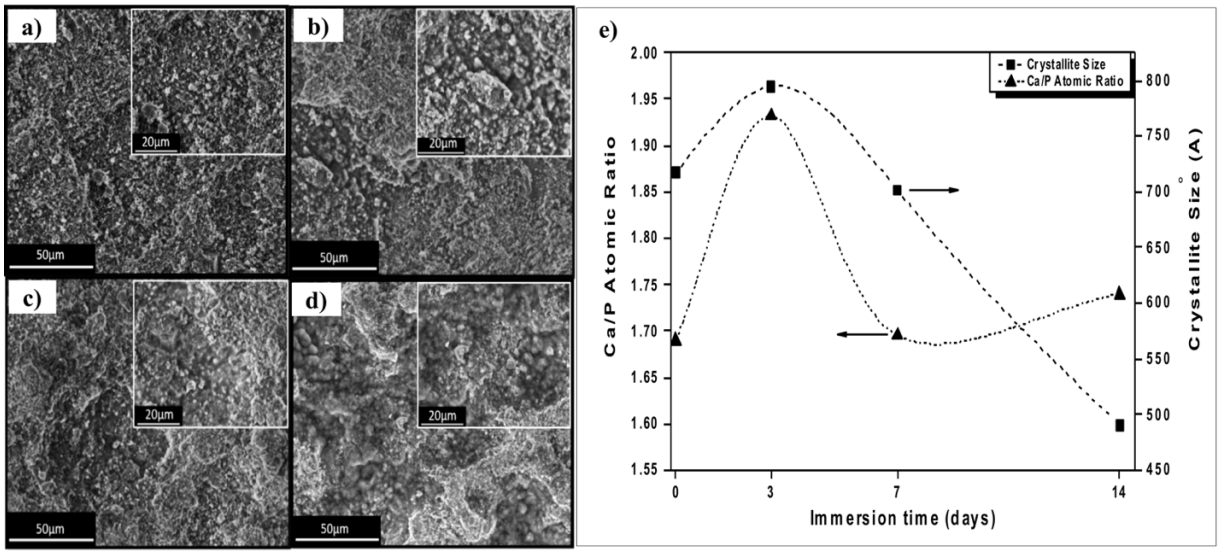

Figure 8.

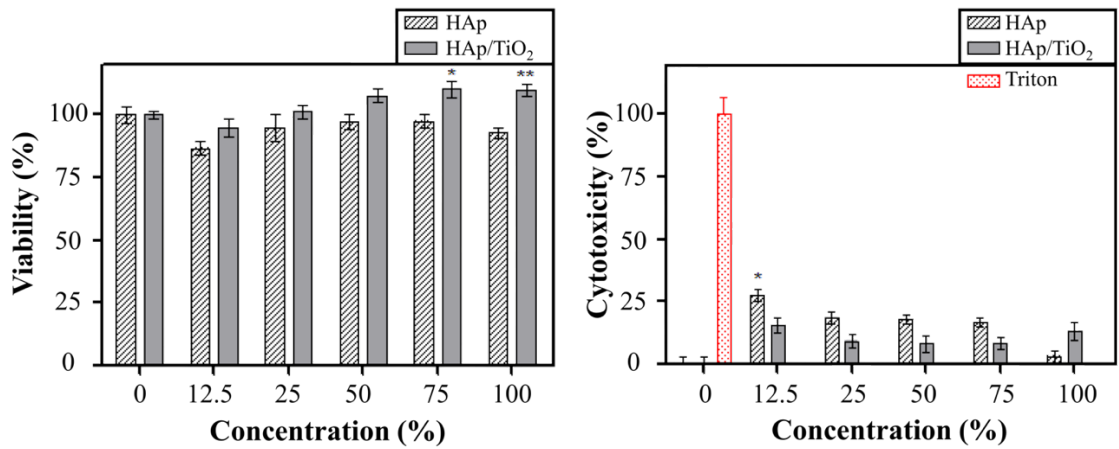

Figure 9.